\documentclass[preprint,aps,fleqn]{revtex4}
\usepackage{graphicx}
\usepackage{dcolumn}

\begin{document}

\title{Determining the antiproton magnetic moment from measurements of the
hyperfine structure of antiprotonic helium}

 \author{Dimitar Bakalov}
 \affiliation{Institute for Nuclear Research and Nuclear Energy,
 Tsarigradsko chauss\'{e}e 72, Sofia 1784, Bulgaria}
 \author{Eberhard Widmann}
 \affiliation{Stefan Meyer Institute, Austrian Academy of Sciences,
Boltzmanngasse 3, 1090 Vienna, Austria}

\begin{abstract}
 Recent progress in the spectroscopy of antiprotonic helium has
 allowed for measuring the separation between components of the
 hyperfine structure (HFS) of the $(37,35)$ metastable states
 with an accuracy of 300 kHz, equivalent to
 a relative accuracy of $3.10^{-5}$. The analysis of the uncertainties
 of the available theoretical results on the antiprotonic helium HFS
 shows that the accuracy of the value of the dipole magnetic moment of
 the antiproton
 (currently known to only 0.3\%) may be improved by up to 2 orders
 of magnitude by measuring the splitting of appropriately
 selected components of the HFS of any of the known metastable states.
 The feasibility of the proposed measurement by means of an
 analog of the triple resonance method is also discussed.
 \end{abstract}
 \maketitle

\section{Introduction}

 Precision spectroscopy of antiprotonic helium is among the most
 spectacular examples of a successful fusion of
 particle accelerator with low energy atomic physics
 methods for the study of the fundamental characteristics of
 an elementary particle - the antiproton
 (see Refs.~\cite{revmodphys,physrep} and references therein.)
 %
 Among the main goals of the experimental program of the CERN
 collaborations PS205 and ASACUSA are
 precision tests of bound-states QED, the determination of the
 dipole magnetic moment of the antiproton, and independent tests of CPT
 invariance.
 Strong limits of {\bf $2\times10^{-9}$} on
 the possible differences between proton and antiproton masses and
 electric charges have already been extracted from
 the experimental results \cite{hori-prl2006}.
 Studies of QED of  bound systems involving antiparticles
 are motivated by the unsolved problems \cite{karsh-pos}
 in the theoretical evaluation of the hyperfine structure (HFS) of
 positronium \cite{pos-th}, which is known not to be in perfect
 agreement with experiment \cite{pos-exp}. It is believed that
 QED tests on systems involving heavy antiparticles 
 may help
 understand these problems better, since the various QED contributions
 have different weights in antiprotonic helium as compared to
 positronium.
 In the present paper we focus our attention on the possibility of
 determining the antiproton magnetic moment (currently known to
 0.3\% only  from a measurement of the fine structure of antiprotonic lead 
  \cite{PDG}) with an improved
 accuracy by measuring the hyperfine splitting and comparing
 the spectroscopy data with the
 theoretical calculations of the hyperfine structure (HFS) of
 metastable states of the ${\bar{p}}^4\!${\rm He}\
 atoms \cite{HFS98,HFS01}. While the new value
 will be too much less accurate than the value
 of the magnetic moment of the proton \cite{mohr05} for a
 meaningful test of CPT, it will fill a blank in the particle
 properties tables that has survived
 for more than 2 decades.

 In the non-relativistic approximation the bound states of
 ${\bar{p}}^4\!${\rm He}\
 are traditionally labelled with
 the quantum numbers of the total orbital momentum $L$ and the
 principal quantum number $n$, though an alternative labelling
 with $L$ and the vibrational quantum number $v$ is also used; of
 course, $n=L+v+1$.
 For the near-circular excited states with $L$ in the range $L\geq30$ and
 small $v$ the Auger decay is suppressed (Condo
 mechanism \cite{condo}) so that they de-excite only through
 slow radiative transitions.
 The life time of these states may reach microseconds;
 they are referred to as metastable.

\begin{figure}[ht]
 \includegraphics{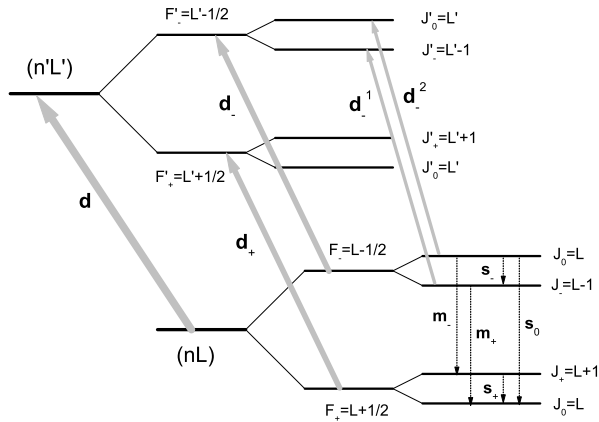}
 \caption{Hyperfine structure of a pair of parent and daughter states
 of ${\bar{p}}^4\!${\rm He}\
 and of the dipole transition $(nL)\rightarrow(n'L')$ between them.
 The transitions between states of the $F_{-}$ and  $F_{+}$
 doublets are denoted
 by $m_-$, $m_+$ and $m_0$ depending on $\Delta J$; the transitions
 within the $F_{\pm}$ doublets are labelled as $s_{\pm}$.
 The transitions between homologous doublets of the parent and
 daughter states are denoted by $d_{\pm}$, and between the
 homologous components of the doublets - by $d_{\pm}^{1,2}$.}
 \label{hfs}
 \end{figure}

 The pairwise spin interactions between the constituents
 of ${\bar{p}}^4\!${\rm He}\
 split each Coulomb level into hyperfine components \cite{HFS98,HFS01}.
 The hyperfine structure of the metastable state $(nL)$ consists of 4
 substates $(nLFJ)$, labelled (in addition to $n$ and $L$) with the
 quantum numbers $F$ and $J$ of the intermediate angular momentum
 $\mathbf{F}=\mathbf{L}+\mathbf{s}_e$ and the total angular momentum
 $\mathbf{J}=\mathbf{F}+\mathbf{s}_{\bar{p}}$;
 here $\mathbf{s}_e$ and $\mathbf{s}_{\bar{p}}$ stand for the
 spin operators of the electron and the antiproton. The spin
 interactions are dominated by the electron spin-orbit
 interaction causing a splitting of the order of 10 GHz of the
 $(nL)$ level into the $F_{\pm}$ doublets with
 $F=L\pm1/2$. The splitting within the $F_{\pm}$
 doublets is due to interactions involving the antiproton spin,
 and is approximately two orders of magnitude smaller (see Fig.\ref{hfs}).

 The HFS of the $(37,35)$ state was first observed
 in 1997 \cite{firstobs}, when improved resolution
 allowed for clearly
 distinguishing two peaks in the profile of the
 $(37,35)\rightarrow(38,34)$ transition line.
 The peaks correspond
 to the $d_{-}$ and $d_{+}$ transitions on Fig.\ref{hfs}; at that
 time the components $d_{\pm}^{1,2}$ could not be resolved, and the
 remaining non-diagonal components were too strongly suppressed
 to be observed \cite{HFS98}.
 The first laser spectroscopy study of the HFS of the
 $(37,35)$ state was
 performed in 2002 \cite{wid-hfs}; using the triple resonance
 method the frequencies of the $m_{\pm}$ transitions were measured
 with an accuracy of the order of 300 kHz (below 30 ppm). The idea
 of the method was to depopulate the $F_{-}$ doublet with a laser
 pulse tuned at the frequency of the $d_{-}$ transition, then
 refill it from the $F_{+}$ doublet by applying an oscillating magnetic
 field tuned at the $m_{-}$ or $m_{+}$ transition frequency, and
 measure the refilling rate with a repeated laser pulse tuned at the $d_{-}$
 transition frequency.
 In future measurements of the $m_{\pm}$ transition frequencies
 the experimental uncertainty is expected to be further reduced.
 In what follows we are analyzing the restrictions that
 theoretical and experimental uncertainties impose on the value of
 the antiproton dipole magnetic moment as extracted from
 spectroscopy data, and outline an alternative approach to
 improving the accuracy of this value, possibly by up to 2 orders
 of magnitude.

 \section{Hyperfine structure of the energy levels of
 the metastable states of ${}^{4\!}$\protect{\rm \bf He}}
 \label{3body}

 The spin interaction Hamiltonian $V$, used in the calculations of
 the HFS of ${\bar{p}}^4\!${\rm He}\
 in \cite{HFS98,HFS01}, has the form of a sum of
 pairwise interaction terms:
 $V=V_{\alpha e}+V_{\alpha \bar{p}}+V_{\bar{p}e}$, with
 (in units $\hbar=e=1$)
 \begin{eqnarray}
 &&V_{\alpha e} =  \alpha^2 \left\{ \right.
 \left( 1+2\mu_{e} \right) {1\over m^2_{e}r_{\alpha e}^3}
 (\mathbf{r}_{\alpha e}
 \times \mathbf{p}_{e}) \cdot \mathbf{s}_{e}- {2\mu_{e} \over
 m_{e}m_{\alpha}r_{\alpha e}^{3}} (\mathbf{r}_{\alpha e}
 \times \mathbf{p}_{\alpha}) \cdot
 \mathbf{s}_{e} \left. \right\} \label{Vae}
 \\
 &&V_{\alpha \bar{p}} =  \alpha^2 \left\{ \right.
 \left( 1+2\mu_{\bar{p}} \right) {1\over m^2_{\bar{p}}
 r_{\alpha \bar{p}}^3} (\mathbf{r}_{\alpha \bar{p}}
 \times \mathbf{p}_{\bar{p}}) \cdot \mathbf{s}_{\bar{p}}- {2\mu_{\bar{p}} \over
 m_{\bar{p}}m_{\alpha}r_{\alpha \bar{p}}^{3}} (\mathbf{r}_{\alpha \bar{p}}
 \times \mathbf{p}_{\alpha}) \cdot
 \mathbf{s}_{\bar{p}} \left. \right\}  \label{Vap}
 \\
 &&V_{\bar{p}e} = \alpha^2 \left\{ \right.
 -\frac{8\pi}{3}\, \frac{\mu_{\bar{p}}\,\mu_e}{m_{\bar{p}}\,m_e}
 (\mathbf{s}_{\bar{p}}\cdot\mathbf{s}_{e})\,\delta(\mathbf{r}_{\bar{p}e})
 -{1 \over r^{5}_{\bar{p}e}} \frac{\mu_{\bar{p}}\,\mu_e}{m_{\bar{p}}\,m_e}
 \left(
 3(\mathbf{r}_{\bar{p}e}\cdot\mathbf{s}_{\bar{p}})
 (\mathbf{r}_{\bar{p}e}\cdot\mathbf{s}_{e})-
 r_{\bar{p}e}^2(\mathbf{s}_{\bar{p}}\cdot\mathbf{s}_{e}) \right)
 \nonumber \\
 & & -\left( 1+2\mu_{\bar{p}} \right) {1 \over 2m^2_{\bar{p}}r_{\bar{p}e}^3}
 (\mathbf{r}_{\bar{p}e} \times \mathbf{p}_{\bar{p}})
 \cdot \mathbf{s}_{\bar{p}}- {\mu_{e}
 \over m_{\bar{p}}m_{e}r_{\bar{p}e}^{3}}
 (\mathbf{r}_{\bar{p}e} \times \mathbf{p}_{\bar{p}})
 \cdot \mathbf{s}_{e} \nonumber \\
 & & +\left( 1+2\mu_{e} \right) {1 \over 2m^2_{e}r_{\bar{p}e}^3}
 (\mathbf{r}_{\bar{p}e} \times \mathbf{p}_{e}) \cdot \mathbf{s}_{e}+ {\mu_{\bar{p}}
 \over m_{\bar{p}}m_{e}r_{\bar{p}e}^{3}} (\mathbf{r} \times \mathbf{p}_{e}) \cdot
 \mathbf{s}_{\bar{p}} \left. \right\}  \label{Vpe}
 \end{eqnarray}
 Here $m_i$, $\mathbf{r}_i$, $\mathbf{p}_i$ and
 $\mathbf{s}_i$, $i=e,\bar{p},\alpha$ stand for the mass,
 position vector, momentum and spin operator of the $i$-th
 constituent of the ${\bar{p}}^4\!${\rm He}\
 atom, $\mathbf{r}_{ij}=\mathbf{r}_{j}-\mathbf{r}_{i}$.
 {\em
 and $\mu_i$ is the magnetic moment of particle $i$ in units
 of ``own magnetons'' $|e_i|\hbar/2m_i\equiv 1/2m_i$,
 $e_i$ being the electric charge in units $e$.
 }
 In first order of perturbation
 theory the hyperfine energy levels $E_{nLFJ}$ are calculated as eigenvalues
 of the matrix of $V$ in an appropriate basis.
 The computational procedure makes use of the effective spin
 Hamiltonian of the system
 -- a finite-dimensional operator acting
 in the space of the spin and orbital momentum variables of the particles:
 \begin{equation}
 H_{\text{eff}}\!=\!H_{1}\,(\mathbf{s}_{e}\cdot\mathbf{L})+ H_{2}\,
 (\mathbf{s}_{\bar{p}}\cdot\mathbf{L})+
 H_{3}\,(\mathbf{s}_{\bar{p}}\cdot\mathbf{s}_{e})+ 
 H_{4} \,( 2L(L+1) (\mathbf{s}_{\bar{p}}\cdot\mathbf{s}_{e})-
 3((\mathbf{s}_{\bar{p}}\cdot\mathbf{L})(\mathbf{s}_{e}\cdot\mathbf{L})+
 (\mathbf{s}_{e}\cdot\mathbf{L})(\mathbf{s}_{\bar{p}}\cdot\mathbf{L}))
 ).
 \label{heff}
 \end{equation}
 The coefficients $H_i, i=1,\ldots,4$ of $H_{\rm eff}$ are
 calculated by averaging the spin interaction Hamiltonian $V$ of
 Eqs.~(\ref{Vae})-(\ref{Vpe}) with the non-relativistic three-body
 wave functions of ${\bar{p}}^4\!${\rm He}; the remaining part of the
 computations is reduced to angular momentum algebra.
 The uncertainty of $V$ is determined by the contribution of the
 interaction terms of order
 $O(m_e\alpha^6)$ and higher, that have not been taken into consideration.
 Accordingly, the relative uncertainty
 $\Delta_{\rm q}H_i$ of the coefficients
 $H_i$, due to truncating the expansion of $V$ in power series in
 $\alpha$ is estimated to be of relative order
 $\Delta_{\rm q}H_i\sim O(\alpha^2)\sim10^{-4}$ \cite{HFS98,HFS01}.
 The uncertainty of $H_i$ gives rise to the uncertainties
 $\delta_{\rm q}E_{nLFJ}$ and
 $\delta_{\rm q}\nu$ of the
 hyperfine energy levels and the hyperfine transition
 frequencies, respectively, and to the relative
 uncertainties
 $\Delta_{\rm q}E_{nLFJ}=\delta_{\rm q}E_{nLFJ}/E_{nLFJ}$
 and $\Delta_{\rm q}\nu=\delta_{\rm q}\nu/\nu$.
 The latter are expressed in terms of the
 response of $E_{nLFJ}$ and $\nu$ to variations of $H_i$
 around the values calculated with the spin interaction
 Hamiltonian $V$,
 and are given by the derivatives
 $R_i(FJ)=
 \partial\Delta_{\rm q}E_{nLFJ}/\partial\Delta_{\rm q}H_i
 \vert_{\Delta_{\rm q}H_i=0}$ and
 $R_i(\nu)=\partial\Delta_{\rm q}\nu/\partial\Delta_{\rm q}H_i
 \vert_{\Delta_{\rm q}H_i=0}$.
 Table~\ref{extra}, presenting the numerical values of these
 derivatives for the hyperfine levels of the $(37,35)$ state, shows
 that the theoretical accuracy for all five allowed hyperfine
 transitions is of the order of $\Delta_{\rm q}H_i\sim10^{-4}$
 since $|R_i|$ does not exceed 1 and no
 precision is lost.
 We have also included in consideration the difference $X$ of the
 transition frequencies of the $m_-$ and $m_+$ transitions,
 $X=\nu(m_-)-\nu(m_+)$. This combination is of interest
 because, on the one hand, it is quite sensitive to the value of
 $\mu_{\bar{p}}$, and on the other, an improvement of
 the precision on the $m_{-}$ and $m_{+}$ transition frequencies
 and therefore also on $X$ by at least one order of magnitude is
 expected in experiments using an improved laser system in the
 near future \cite{ASACUSAprop}.

 \begin{table}[ht]
 \begin{center}
 \caption{Response of the hyperfine energy levels $E_{nLFJ}$,
 of the hyperfine transition frequencies $\nu$ and of the
 difference $X$ of the $m_{-}$ and $m_{+}$ transition frequencies
 in the metastable state $(37,35)$ of the ${\bar{p}}^4\!${\rm He}\
 atom to variations
 of the effective spin Hamiltonian coefficients $H_i$.
 Listed are the dimensionless derivatives
 $\partial\Delta_{\rm q}E_{nLFJ}/\partial\Delta_{\rm q}H_i$
 and $\partial\Delta_{\rm q}\nu/\partial\Delta_{\rm q}H_i$,
 evaluated numerically at $\Delta_{\rm q}H_i=0$, i.e. using the
 values calculated with the spin interaction Hamiltonian $V$.}
 \label{extra}
 \begin{tabular}{@{\hspace{3mm}}c@{\hspace{3mm}}r@{\hspace{3mm}}r
 @{\hspace{3mm}}r@{\hspace{3mm}}r@{\hspace{3mm}}r@{\hspace{3mm}}
 r@{\hspace{3mm}}r@{\hspace{3mm}}r@{\hspace{3mm}}r@{\hspace{3mm}}
 r@{\hspace{3mm}}}
  \hline\hline
 $i$ & $R(F_{-}J_{0})$ & $R(F_{-}J_{-})$  &
 $R(F_{+}J_{+})$  & $R(F_{+}J_{0})$ &
 $R(s_{-})$ & $R(s_{+})$ & $R(m_{-})$ & $R(m_{+})$ & $R(m_{0})$
 & \multicolumn{1}{c}{$R(X)$} \\
 \hline
  1 & 1.010 & 0.989 & 0.986 & 1.012  & $-$0.031
  & $-$0.025 & 0.998 & 1.000 & 0.988  &
  0.000\\
  2  & $-$0.012 & 0.011 & 0.012  & $-$0.012
  & 1.125 & 0.929  & 0.000  & 0.000 & 0.011
  & 0.000\\
  3  & $-$0.011 & 0.012  & $-$0.011 & 0.012
  & 1.141  & $-$0.903  & $-$0.011 & 0.012 & 0.000  &
  $-$10.613\\
  4 & 0.013  & $-$0.012 & 0.013  & $-$0.012  & $-$1.235
  & 0.999 & 0.013  & $-$0.012 & 0.000 &
  11.613\\
 \hline\hline
 \end{tabular}
 \end{center}
 \end{table}

 Note that the theoretical prediction for $X$
 is less accurate than for the 5 hyperfine transition frequencies.
 The uncertainty of the value of $X$,
 $\Delta_{\rm q}(X)=\max_i\, |R_i(X).\Delta_{\rm q}H_i|$, is
 larger than $10^{-4}$ and is strongly
 state-dependent (see Table~\ref{statedep}).
 The values in the table were calculated with the assumption that the
 uncertainties of $H_i,i=1,\ldots,4$ are not correlated, and
 should be regarded as upper limits for the theoretical
 uncertainties of $X$.

 \begin{table}[ht]
 \begin{center}
 \caption{Relative uncertainty $\Delta_{\rm q}(X)$ of the
 theoretical value of the frequency
 $X$ in the metastable states of the ${\bar{p}}^4\!${\rm He}\ atom,
 due to neglecting the higher-order terms in the spin interaction
 Hamiltonian $V$ of Eqs.~(\ref{Vae})-(\ref{Vpe}).}
 \label{statedep}
 \begin{tabular}{@{\hspace{3mm}}c@{\hspace{3mm}}c@{\hspace{3mm}}c
 @{\hspace{3mm}}c@{\hspace{3mm}}c@{\hspace{3mm}}c@{\hspace{3mm}}
 c@{\hspace{3mm}}c@{\hspace{3mm}}c@{\hspace{3mm}}c@{\hspace{3mm}}}
  \hline\hline
 $(n,L)$ & $(35,33)$ & $(37,34)$ & $(39,35)$ & $(33,32)$ &
 $(36,34)$ & $(37,35)$ & $(35,34)$ & $(34,33)$ & $(38,35)$ \\
 $\Delta_{\rm q}(X)$ & $6\times10^{-4}$ & $11\times10^{-4}$
 & $3\times10^{-4}$ & $8\times10^{-4}$
 & $23\times10^{-4}$ & $12\times10^{-4}$
 & $6\times10^{-4}$ & $4\times10^{-4}$
 & $5\times10^{-4}$\\
 \hline\hline
 \end{tabular}
 \end{center}
 \end{table}

 The dominating contribution to $E_{nLFJ}$
 comes from the electron spin--orbit interaction
 which does not depend of the value of the dipole magnetic moment of
 the antiproton.
 The value of $\mu_{\bar{p}}$ may be determined from spectroscopy
 data about the HFS of ${\bar{p}}^4\!${\rm He}\ if one
 selects hyperfine transitions whose frequencies
 depend as strongly as possible
 on the value of $\mu_{\bar{p}}$. To help making the appropriate choice,
 we calculate - for the 9 metastable
 states that have been experimentally observed by now -
 the ``sensitivity''
 of the hyperfine levels $E_{nLFJ}$ and of
 the transition frequencies between them
 to variations of $\mu_{\bar{p}}$ around the CPT-prescribed value
 $\mu_{\bar{p}}=-\mu_p$.
 (In agreement with \cite{gabrielse} we neglect the effects of
 the small difference of less than $10^{-10}$
 between the ``own magnetons '' of the proton and
 antiproton). We define
 the sensitivity $S(FJ)\equiv S(nLFJ)$ of the hyperfine level $E_{nLFJ}$ as:
 \begin{equation}
 S(FJ)=\partial E_{nLFJ}/\partial \mu_{\bar{p}}
 \vert_{\mu_{\bar{p}}=-\mu_p}
 \end{equation}
 The sensitivity of a transition frequency is then the
 difference of the sensitivities of the initial and final states,
 e.g. $S(s_{-})=S(F_{-}J_{0})-S(F_{-}J_{-})$,
 $S(m_{0})=S(F_{-}J_{0})-S(F_{+}J_{0})$, etc.
 The sensitivity values in Table~\ref{sens} have been calculated by
 numerical differentiation
 of the eigenvalues $E_{nLFJ}$ of the spin
 interaction Hamiltonian.
 Because of the opposite signs of $S$ for the upper and lower
 sublevels in the $F_{-}$ and $F_{+}$ doublets, the sensitivity of
 the $s_{-}$, $s_{+}$ and $m_{0}$ hyperfine transitions is
 enhanced, while the sensitivity of the $m_{-}$ and $m_{+}$
 transitions is
 suppressed by orders(s) of magnitude (see Fig.~\ref{hfs}).

 \begin{table}[ht]
 \begin{center}
 \caption{Sensitivities $S(FJ)$ of the hyperfine
 sublevels from the HFS of
 a selection of metastable states of the ${\bar{p}}^4\!${\rm He}\ atom
 to variations of the magnetic moment of the antiproton,
 and sensitivities $S$ of the
 hyperfine transitions between these sublevels (see Fig.~\ref{hfs})
 and of the difference $X=\nu(m_-)-\nu(m_+)$,
 in units MHz.}
 %
 \label{sens}
 \begin{tabular}{@{\hspace{3mm}}c@{\hspace{3mm}}r@{\hspace{3mm}}r
 @{\hspace{3mm}}r@{\hspace{3mm}}r@{\hspace{3mm}}r@{\hspace{3mm}}
 r@{\hspace{3mm}}r@{\hspace{3mm}}r@{\hspace{3mm}}r@{\hspace{3mm}}
 r@{\hspace{3mm}}}
  \hline
  \hline
 $(nL)$ & $S(F_{-}J_{0})$ & $S(F_{-}J_{-})$  &
 $S(F_{+}J_{+})$  & $S(F_{+}J_{0})$ &
 $S(s_{-})$ & $S(s_{+})$ & $S(m_{-})$ & $S(m_{+})$ & $S(m_{0})$ & $S(X)$\\
 \hline
 $(35,33)$  & $-$52.6{\hspace{3mm}} &  54.8{\hspace{3mm}} & $-$45.5{\hspace{3mm}}
 &  40.5{\hspace{3mm}} & $-$107.4{\hspace{1mm}} &  $-$86.0{\hspace{1mm}}
 &  $-$7.1{\hspace{2mm}} &  14.2{\hspace{2mm}} & 100.2{\hspace{1mm}}
 & $-$21.3{\hspace{4mm}}\\
 $(37,34)$  & $-$30.6{\hspace{3mm}} &  32.6{\hspace{3mm}} & $-$38.9{\hspace{3mm}}
 &  34.9{\hspace{3mm}} &  $-$63.2{\hspace{1mm}} &  $-$73.8{\hspace{1mm}}
 &   8.4{\hspace{2mm}} &  $-$2.3{\hspace{2mm}} & 71.5{\hspace{1mm}}
 & 10.7{\hspace{4mm}}\\
 $(39,35)$  & $-$15.0{\hspace{3mm}} &  16.9{\hspace{3mm}} & $-$33.7{\hspace{3mm}}
 &  30.6{\hspace{3mm}} &  $-$31.9{\hspace{1mm}} &  $-$64.4{\hspace{1mm}}
 &  18.8{\hspace{2mm}} & $-$13.7{\hspace{2mm}} & 50.6{\hspace{1mm}}
 & 32.5{\hspace{4mm}}\\
 $(33,32)$  &    $-$81.8{\hspace{3mm}} &     83.9{\hspace{3mm}}
 &    $-$55.3{\hspace{3mm}} &     49.2{\hspace{3mm}} &    $-$165.6{\hspace{1mm}}
 &    $-$104.5{\hspace{1mm}} &    $-$26.4{\hspace{2mm}} &     34.6{\hspace{2mm}}
 &    139.2{\hspace{1mm}}
 & $-$61.1{\hspace{4mm}}\\
 $(36,34)$  & $-$39.5{\hspace{3mm}} &  41.7{\hspace{3mm}} & $-$40.2{\hspace{3mm}}
 &  35.8{\hspace{3mm}} &  $-$81.2{\hspace{1mm}} &  $-$76.0{\hspace{1mm}}
 &   0.7{\hspace{2mm}} &   5.8{\hspace{2mm}} & 81.8{\hspace{1mm}}
 & $-$5.1{\hspace{4mm}}\\
 $(37,35)$  & $-$28.2{\hspace{3mm}} &  30.3{\hspace{3mm}} & $-$36.2{\hspace{3mm}}
 &  32.3{\hspace{3mm}} &  $-$58.5{\hspace{1mm}} &  $-$68.5{\hspace{1mm}}
 &   8.0{\hspace{2mm}} &  $-$2.1{\hspace{2mm}} & 66.5{\hspace{1mm}}
 & 10.1{\hspace{4mm}}\\
  $(35,34)$  & $-$50.1{\hspace{3mm}} &  52.3{\hspace{3mm}} & $-$41.3{\hspace{3mm}}
 &  36.5{\hspace{3mm}} & $-$102.4{\hspace{1mm}} &  $-$77.9{\hspace{1mm}}
 & $-$8.7{\hspace{2mm}} & 15.8{\hspace{2mm}} & 93.7{\hspace{1mm}}
 & $-24.5${\hspace{4mm}}\\
 $(34,33)$  & $-$64.8{\hspace{3mm}} &  67.0{\hspace{3mm}} & $-$47.3{\hspace{3mm}}
 &  41.9{\hspace{3mm}} & $-$131.9{\hspace{1mm}} &  $-$89.2{\hspace{1mm}}
 & $-$17.5{\hspace{2mm}} & 25.1{\hspace{2mm}} & 114.4{\hspace{1mm}}
 & $-$42.6{\hspace{4mm}}\\
 $(38,35)$  & $-$20.8{\hspace{3mm}} & 22.8{\hspace{3mm}} & $-$35.1{\hspace{3mm}}
 &  31.7{\hspace{3mm}} &  $-$43.7{\hspace{1mm}} &  $-$66.8{\hspace{1mm}}
 & 14.3{\hspace{2mm}} &     $-$8.8{\hspace{2mm}} &     57.9{\hspace{1mm}}
 & 23.1{\hspace{4mm}}\\
\hline\hline
 \end{tabular}
 \end{center}
 \end{table}

 The current uncertainty in the value of the magnetic moment of
 the antiproton
 $\delta\mu_{\bar{p}}\sim 3.10^{-3}\times\mu_{\bar{p}}\sim
 8.10^{-3}$ 
 gives rise to an uncertainty $\delta_{\mu}\nu$ of
 the theoretical frequency $\nu$ of the various hyperfine
 transitions, that is expressed in terms of the sensitivity $S$:
 $\delta_{\mu}\nu = |S|.\delta\mu_{\bar{p}}$.
 The corresponding relative uncertainty $\Delta_{\mu}\nu$ is given
 by $\Delta_{\mu}\nu=\delta_{\mu}\nu/\nu$.
 A measurement of the frequency $\nu$ of a hyperfine transition
 with an experimental uncertainty $\delta_{\rm exp}\nu$
 could improve the current accuracy of the antiprotonic magnetic moment
 value only if (1) the experimental error is sufficiently smaller
 than the theoretical uncertainties
 $\delta_{\mu}\nu$ and $\delta_{\rm q}\nu$, and (2)
 $\delta_{\rm q}\nu < \delta_{\mu}\nu$ or, equivalently,
 $\Delta_{\mu}\nu/\Delta_{\rm q}\nu>1$.
 Table~\ref{uncert} presents the value of
 the absolute uncertainty $|\delta_{\mu}\nu|$
 and of the ratio $|\Delta_{\mu}\nu/\Delta\nu_{\rm q}|$
 for all hyperfine transitions in the
 nine observed metastable states of the ${\bar{p}}^4\!${\rm He}\ atom.
 In absence of more precise theoretical calculation which
 take consistently into account all QED and
 relativistic effects of order $O(m_e\alpha^6)$, we have
 assumed (in agreement with the results in Table \ref{extra}) that
 $\Delta_{\rm q}\nu=10^{-4}$ for all hyperfine transitions.
 For the difference $X$ of the $m_{-}$ and $m_{+}$ transition
 frequencies we used the values of  $\Delta_{\rm q}(X)$
 from Table~\ref{statedep}.

 \begin{table}[ht]
 \begin{center}
 \caption{Absolute uncertainty $\delta_{\mu}\nu$ (in MHz),
 related to the current uncertainty of 0.3\%
 of the magnetic dipole moment of the antiproton, and
 the ratio $\Delta_{\mu}\nu/\Delta_{q}\nu$ of the relative
 theoretical uncertainties $\Delta_{\mu}\nu$ and $\Delta_{q}\nu$ of
 the hyperfine transition frequencies in the metastable
 states $(nL)$ of ${\bar{p}}^4\!${\rm He}.
 For the labelling of the hyperfine transitions, see
 Fig.~\ref{hfs}; $X$ labels the difference of the $m_{-}$ and
 $m_{+}$ transition frequencies.}
 \label{uncert}
 \begin{tabular}{c@{\hspace{3mm}}l@{\hspace{3mm}}dddddd}
  \hline\hline
 \vrule width 0pt height 10pt
 $(nL)$ & &
 \multicolumn{1}{r}{$s_{-}$} &
 \multicolumn{1}{r}{$s_{+}$} &
 \multicolumn{1}{r}{$m_{-}$} &
 \multicolumn{1}{r}{$m_{+}$} &
 \multicolumn{1}{c}{$m_{0}$} &
 \multicolumn{1}{c}{$X$} \\
 \hline
 \vrule width 0pt height 10pt
 $(35,33)$  & $\delta_{\mu}\nu$ & 0.90 &  0.72
 &  0.06 &  0.12 & 0.84 & 0.18\\
 & $\Delta_{\mu}\nu/\Delta_{\rm q}\nu$ & 35.3 &  36.9
 &  0.0 &  0.1 & 0.6 & 5.0\\
 \vrule width 0pt height 15pt
 %
 $(37,34)$  & $\delta_{\mu}\nu$ &  0.53 &  0.62
 &   0.07 &  0.02 & 0.60 & 0.09\\
 & $\Delta_{\mu}\nu/\Delta_{\rm q}\nu$ &  36.8 &  35.6
 &   0.1 &  0.0 & 0.6 & 2.7\\
 \vrule width 0pt height 15pt
 %
 $(39,35)$  & $\delta_{\mu}\nu$ &  0.27 &  0.54
 &  0.16 & 0.11 & 0.42 & 0.27\\
 & $\Delta_{\mu}\nu/\Delta_{\rm q}\nu$ &  40.6 &  34.4
 &  0.1 & 0.1 & 0.4 & 8.9\\
 \vrule width 0pt height 15pt
 %
 $(33,32)$  & $\delta_{\mu}\nu$ &    1.39 &    0.88 &   0.22
 &    0.29 &    1.17 & 0.51\\
 & $\Delta_{\mu}\nu/\Delta_{\rm q}\nu$ &   34.6 & 38.0 & 0.1 & 0.2 & 0.8 & 3.6\\
 \vrule width 0pt height 15pt
 %
 $(36,34)$  & $\delta_{\mu}\nu$ &  0.68 &  0.64
 &   0.00 &   0.05 & 0.69 & 0.05\\
 & $\Delta_{\mu}\nu/\Delta_{\rm q}\nu$ &  35.9 &  36.3
 &   0.0 &   0.0 & 0.5 & 1.3\\
 \vrule width 0pt height 15pt
 %
 $(37,35)$  & $\delta_{\mu}\nu$ &  0.49 &  0.57
 &   0.07 &  0.02 & 0.56 & 0.09\\
 & $\Delta_{\mu}\nu/\Delta_{\rm q}\nu$ &  36.9 &  35.7
 &   0.1 &  0.0 & 0.4 & 2.7\\
 %
 \vrule width 0pt height 15pt
 %
 $(35,34)$  & $\delta_{\mu}\nu$ &  0.86 &  0.65
 &   0.07 &  0.13 & 0.79 & 0.21\\
 & $\Delta_{\mu}\nu/\Delta_{\rm q}\nu$ &  35.3 &  37.1
 &   0.0 &  0.1 & 0.5 & 5.4\\
 \vrule width 0pt height 15pt
 %
 $(34,33)$  & $\delta_{\mu}\nu$ &  1.11 &  0.75
 &   0.15 &  0.21 & 0.96 & 0.36\\
 & $\Delta_{\mu}\nu/\Delta_{\rm q}\nu$ &  35.0 &  37.9
 &   0.1 &  0.1 & 0.6 & 8.4\\
 \vrule width 0pt height 15pt
 %
 $(38,35)$  & $\delta_{\mu}\nu$ &  0.37 &  0.56
 &   0.12 &  0.07 & 0.49 & 0.19\\
 & $\Delta_{\mu}\nu/\Delta_{\rm q}\nu$ &  38.8 &  35.0
 &   0.1 &  0.1 & 0.4 & 6.0\\
 \hline\hline
 \end{tabular}
 \end{center}
 \end{table}

 To improve the
 current accuracy of 0.3\% of $\mu_{\bar{p}}$, the absolute
 experimental uncertainty $\delta_{\rm exp}\nu$
 of the measurement of the transition frequency $\nu$ should be
 {\em below} the corresponding value $\delta_{\mu}\nu$
 of Table~\ref{uncert}.
 Provided that this condition is fulfilled, the
 ratio $\Delta_{\mu}\nu/\Delta_{\rm q}\nu$ is an estimate of the
 expected factor of improvement of the accuracy of
 $\mu_{\bar{p}}$.
 In other words, the ratio $\Delta_{\mu}\nu/\Delta\nu_{\rm q}$
 is a criterium  for selecting
 the hyperfine transitions that are most appropriate for determining
 $\mu_{\bar{p}}$.
 A quick look at the Table~\ref{uncert} shows that
 measurements of the $s_{-}$ and $s_{+}$
 transitions in any of the metastable states would
 improve the accuracy of the experimental values of
 $\mu_{\bar{p}}$ by a factor between 35 and 40.
 Measurements of the difference $X$ of $m_{-}$ and $m_{+}$
 transition frequencies in the $(39,35)$ and $(34,33)$ states might
 improve the value of $\mu_{\bar{p}}$ by an order of magnitude.
 No gain of accuracy is expected from measurements of
 the $m_{\pm}$ and $m_{0}$ transitions.

 \section{Application of the triple resonance method to measurements
 of the hyperfine transition frequencies}

 The $s_{-}$ and $s_{+}$ transition frequencies
 could be measured using an analog of the triple resonance method
 of Ref.~\cite{wid-hfs}. Initially, the $J_{-}$ and $J_{0}$
 sublevels of the $F_{-}$ doublet (see Fig.~\ref{hfs})
 are equally populated. By applying a laser pulse, tuned at the
 resonance frequency of the $d_{-}^1$ transition and de-tuned from
 the $d_{-}^2$ frequency, the $J_{-}$ and $J_{0}$ sublevels are
 depopulated asymmetrically. Symmetry is (partially) restored by resonance
 magnetic field-stimulated $s_{-}$ transitions. The fulfillment of the resonance
 condition is checked by means of a second, delayed laser pulse of
 the same frequency as the first one, intended to display any increase
 of the population of the $J_{-}$ sublevel. The expected
 difficulties in such a measurement are related to the low
 intensity of the $s_{\pm}$ transition lines and to the overlap of
 the $d_{-}^{1,2}$ transition line profiles that makes the
 efficiency of the asymmetrical depopulation of the $F_{-}$ doublet
 far from obvious.

 The $s_{-}$, $s_{+}$ and $m_{0}$
 transition lines are much weaker than the $m_{-}$ and $m_{+}$
 lines, which were subject to spectroscopy measurements by the
 ASACUSA collaboration in 2002 \cite{wid-hfs}.
 Compared to the Rabi frequency $\nu_R$ of
 $m_{-}$ and $m_{+}$,
 $\nu_R(m_{\pm})\approx (\mu_B B_0)/\sqrt{6}$, the Rabi
 frequencies of $s_{\pm}$ and $m_{0}$ are suppressed by factors of the order
 of $L$:
 $\nu_R(m_{0})/\nu_R(m_{\pm})\sim(1+2\phi L)/L\sqrt{2}$,
 $\nu_R(s_{\pm})/\nu_R(m_{\pm})\sim(1+2\phi L)/2L$,
 where $\phi\sim2.10^{-2}$ is the mixing angle of the $F_{\pm}$ components
 in the $J=L$ hyperfine states (see Table II of Ref.~\cite{HFS98}).
 Precision spectroscopy of the $s_{-}$ and $s_{+}$ transition
 lines would therefore require a longer measurement time and a stronger
 magnetic field, oscillating with frequencies in the 100 -- 200 MHz range.

 \begin{figure}
 \includegraphics{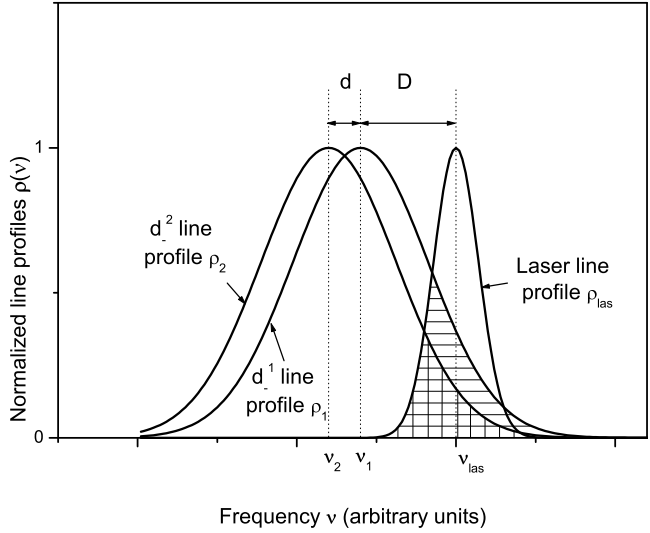}
 \caption{Asymmetric depopulation of the hyperfine states $J_{-}$ and
 $J_{0}$ of the $F_{-}$ doublet.}
 \label{mhz}
 \end{figure}

 To estimate the efficiency of asymmetrical depopulation,
 we consider a simple model in which the laser line profile is
 assumed to be a Gaussian, centered at $\nu_{\rm las}$:
 $\rho_{\rm las}(\nu;\nu_{\rm las},w_{\rm las})=(w_{\rm
 las}\sqrt{\pi})^{-1}
 \exp(-(\nu-\nu_{\rm las})^2/w_{\rm las}^2)$,
 while the profiles
 of the $d_{-}^{i}$ transition lines are assumed to be Voigtian
 (i.e. convolutions of a Gaussian and a Lorenzian),
 centered at $\nu_{i},i\!=\!1,2$
 (see Fig.\ref{mhz}):
 $\rho_{i}(\nu;\nu_{i},w_D,\Gamma_c)=
 (w_D \sqrt{\pi})^{-1}K(\nu/w_D,\Gamma_c/w_D)$,
 where the definition and computational algorithms for the Voigt
 function $K(x,y)$ may be found
 in \cite{voigt3}, $\Gamma_c$ is the collisional HWHM width
 of the $d_{-}^{1,2}$ transition lines, and the parameters
 $w_{\rm las}$ and $w_D$ are related to the FWHM width $\Gamma_{\rm las}$
 of the laser profile and the Doppler width $\Gamma_D$ of
 the $d_{-}^{1,2}$ lines by means of
 $w_{\rm las}=\Gamma_{\rm las}/2\sqrt{\,\log2}$, and similar for
 $w_{D}$.
 The depopulation rates of the $J_{-}$ and $J_{0}$ sublevels,
 $\lambda_{1,2}$, are
 proportional to the overlap of the laser line profile
 with the profiles of the $d_{-}^{1,2}$ transition lines:
 $\lambda_{i}(\nu_i\!-\!\nu_{\rm las})={\rm const}.\int d\nu\,
 \rho_{\rm las}(\nu;\nu_{\rm las},w_{\rm las})\,
 \rho_{i}(\nu;\nu_{i},w_d,\Gamma_c)$, where the dependence on the
 detuning $(\nu_i-\nu_{\rm las})$ has been displayed explicitly.
 Denote by $d$ the distance between the $d_{-}^1$ and $d_{-}^2$
 transition frequencies: $d=\nu_1-\nu_2$, and by $D=(\nu_{\rm las}-\nu_1)$
 -- the detuning between $\nu_1$ and $\nu_{\rm las}$ (see
 Fig.~\ref{mhz}).
 The depopulation asymmetry is described with the
 ratio of the rates of depopulation of the
 $J_{-}$ and $J_{0}$ states $q(D)=\lambda_1(D)/\lambda_2(D+d)$.
 The depopulation rate
 $\lambda_{1}$ may be arranged to exceed
 $\lambda_{2}$ by the factor
 $q>1$ by choosing the detuning $D$ to satisfy the
 nonlinear equation $q(D)=q$.
 This equation has real solutions only for $q$ in the range $1\le
 q\le q_{\rm max}$, with different $q_{\rm max}$ for each transition
 depending on the values of $\Gamma_c$,
 $\Gamma_d$ and $\Gamma_{\rm las}$.
 The price for the achieved asymmetry will be a
 smaller overlap of the $\rho_{\rm las}$ and $\rho_1$ profiles, and,
 as a consequence -- waste of laser power and lower
 $\lambda_{1}$ rate. The waste of laser power may be described
 in terms of the ``power loss factor'' $f=f(D)$, defined as
 \begin{equation}
 f\!\equiv\!f(D)=\!
 \int d\nu\,\rho_{1}(\nu;0,w_d,\Gamma_c)\,\rho_{\rm las}(\nu;D,w_{\rm las})\,\big/
 \!\int d\nu\,\rho_{1}(\nu;0,w_d,\Gamma_c)\,\rho_{\rm las}(\nu;0,w_{\rm
 las}).
 \end{equation}
 To get a quantitative idea of the discussed phenomena,
 we calculate -- for all ten transitions in consideration --
 the values of the detuning $D$ that lead to asymmetrical
 depopulation rates ratio $q=1.2$ and $q=1.5$ (if
 these values are within the range $[1,q_{\rm max}]$), as well as
 the related power loss factor $f$, using
 the realistic value 100 MHz for the FWHM of the laser profile
 \cite{current}. The collisional HWHM
 widths $\Gamma_c$ were calculated for temperature $T=6^{\circ}K$
 and helium gas target number density $3\times10^{20}\ cm^{-3}$
 using the results of  Ref.~\cite{prl2k}.
 The numerical results are presented in Table \ref{res}.

 \begin{table}[ht]
 \begin{center}
 \caption{Values of the detuning $D$ (in MHz)
 of the frequency $\nu_{\rm las}$ of the depopulating laser,
 reckoned from the frequency $\nu_1$
 of the dipole transitions $d^1_-$ from the $F_{-}$ hyperfine
 doublet, that lead to asymmetric depopulation of the
 doublet hyperfine states with a depopulation rate ratio $q=120$\%
 and $q=150$\%.
 The space is left empty when $q$ exceeds the maximal accessible
 values $q_{\rm max}$ for the transition.
 Also listed are the associates laser power loss
 factors $f$.}
 \label{res}
 \begin{tabular}{@{\hspace{3mm}}c@{\hspace{3mm}}r@{\hspace{3mm}}r
 @{\hspace{7mm}}r@{\hspace{7mm}}r@{\hspace{6mm}}r@{\hspace{3mm}}r@{\hspace{6mm}}
 r@{\hspace{3mm}}r@{\hspace{6mm}}r@{\hspace{3mm}}}
  \hline\hline
 \multicolumn{6}{c}{asymmetric depopulation rate ratio $q$} &
 \multicolumn{2}{c}{120\%}
 & \multicolumn{2}{c}{150\%}\\ \hline
 $(nL)\rightarrow(n'L')$ & $\lambda$ (nm) & $\Gamma_c$ (MHz)\hspace*{-4mm} &
 $\Gamma_d$ (MHz)\hspace*{-4mm} & $d$ (MHz)\hspace*{-3mm} & $q_{\rm max}$ &
 $D$ (MHz)\hspace*{-3mm} & \multicolumn{1}{c}{$f$} & $D$ (MHz)\hspace*{-3mm} &
 \multicolumn{1}{c}{$f$}
 \\
 \hline
 $(39,35)\rightarrow(38,34)$ &597 & 108 & 393 & 40.9 & 1.26 & 218 & 0.62\\
 $(37,34)\rightarrow(36,33)$ &470 & 24 & 499 & 57.1 & 1.62 & 136 & 0.84 &
 359 & 0.29 \\
 $(35,33)\rightarrow(34,32)$ &372 & 9 & 630 & 75.1 & 1.92 & 146 & 0.87 &
 375 & 0.39 \\
 $(33,32)\rightarrow(32,31)$ &296 & 6 & 792 & 77.9 & 1.81 & 234 & 0.79 &
 575 & 0.24 \\
 $(37,35)\rightarrow(38,34)$ &726 &  75 & 323 & 26.3 & 1.21 & 247 & 0.40\\
 $(36,34)\rightarrow(37,33)$ &617 & 33 & 380 & 33.9 & 1.37 & 163 & 0.66\\
 $(37,34)\rightarrow(38,33)$ &714 & 90 & 328 & 25.0 & 1.18\\
 %
 $(35,34)\rightarrow(36,33)$ &533 & 15 & 440 & 42.5 & 1.55 & 148 & 0.76 &
 387 & 0.16 \\
 $(34,33)\rightarrow(35,32)$ &458 & 9 & 512 & 48.5 & 1.54 & 167 & 0.76 &
 416 & 0.18 \\
 $(38,35)\rightarrow(39,34)$ &842 & 183 & 278 & 18.5 & 1.09 \\
 %
 \hline\hline
 \end{tabular}
 \end{center}
 \end{table}

 \section{Conclusions}

 We have shown that high accuracy measurements of appropriate
 hyperfine transition lines in the metastable states of
 antiprotonic helium can help reduce the experimental uncertainty
 of the dipole magnetic moment of the antiproton.
 An improvement of the current experiment measuring the $m_+$,
 $m_-$ and as a consequence $X$, is being prepared and it is
 expected to improve the accuracy on  $\mu_{\bar{p}}$ by up to a
 factor of 9. A larger improvement by a factor of up to 40 is possible
 by directly measuring the antiproton spin-flip transitions $s_+$ and
 $s_-$.
 The restrictions on the expected gain of accuracy come from
 the difficulty to reduce the experimental uncertainty below
 the threshold $\delta_{\mu}\nu$ in Table~\ref{uncert}
 rather than from the limited accuracy of the Breit spin interaction
 Hamiltonian $V$ of Eqs.~(\ref{Vae})-(\ref{Vpe}) used in the theoretical
 calculations. We have also outlined a possible experimental
 method for the measurement of the super-hyperfine splitting,
 without discussing in details the feasibility of the experiment.
 We leave for future works the numerical simulations that will
 answer questions about the restrictions on the experimental
 accuracy from the expectedly rather low signal-to-noise ratio and
 about the possible use of a large oscillatory magnetic filed in
 cryogenic helium gas target.

 The authors express their gratitude to Dr. V.I.Korobov for the many
 fruitful discussions on the subject.

\end{document}